\documentclass[journal,singleside,final]{IEEEtran}
\usepackage[nocompress]{cite}
\usepackage{amsmath}
\usepackage{amsthm}
\usepackage{amsfonts}
\usepackage{amssymb}
\usepackage[linesnumbered,ruled,vlined]{algorithm2e}
\usepackage{graphicx}
\usepackage{color}
\usepackage[dvipsnames]{xcolor}
\usepackage[hidelinks]{hyperref}
\usepackage[tight]{subfigure}
\usepackage{enumerate}
\usepackage{multirow}
\definecolor{orange}{rgb}{1,0.5,0}
\definecolor{grn}{rgb}{0.16, 0.5, 0.0}
\makeatletter
\newcommand{\removelatexerror}{\let\@latex@error\@gobble}
\makeatother

\SetKwInput{KwInput}{Input}                
\SetKwInput{KwOutput}{Output}              
\SetKw{Continue}{continue}
\SetKw{Break}{break}

\hypersetup{
    colorlinks=true,
    citecolor=blue,
    linkcolor=black,
    filecolor=magenta,      
    urlcolor=black,
    pdftitle={Overleaf Example},
    pdfpagemode=FullScreen,
    }

\usepackage[acronym,nonumberlist,style=super,toc]{glossaries}

\makeatletter
\newcommand*{\glsplainhyperlink}[2]{%
  \colorlet{currenttext}{.}
  \colorlet{currentlink}{\@linkcolor}
  \hypersetup{linkcolor=currenttext}
  \hyperlink{#1}{#2}%
  \hypersetup{linkcolor=currentlink}
}
\let\@glslink\glsplainhyperlink
\makeatother

\newacronym{5g}{5G}{fifith generation}
\newacronym{6g}{6G}{sixth generation}

\newacronym{awgn}{AWGN}{additive white Gaussian noise}
\newacronym{ao}{AO}{alternating optimization}

\newacronym{bs}{BS}{base station}
\newacronym{bdris}{BD-RIS}{beyond diagonal reconfigurable intelligent surfaces}

\newacronym{csi}{CSI}{channel state information}
\newacronym{cdf}{CDF}{cumulative  distribution  function}

\newacronym{dl}{DL}{downlink}
\newacronym{dris}{D-RIS}{diagonal reconfigurable intelligent surfaces}

\newacronym{fc}{FC}{fully-connected}

\newacronym{gc}{GC}{group-connected}

\newacronym{iid}{i.i.d}{independent and identically distributed}
\newacronym{iot}{IoT}{Internet of Things}
\newacronym{irs}{IRS}{intelligent reflecting surfaces}
\newacronym{isac}{ISAC}{integrated sensing and communication}

\newacronym{jmld}{JMLD}{joint-multiuser maximum likelihood detector}

\newacronym{lut}{LUT}{look-up table}
\newacronym{lhs}{LHS}{left-hand side}

\newacronym{mpsk}{$M$-PSK}{M-ary phase shift keying}
\newacronym{mimo}{MIMO}{multiple-input-multiple-output}
\newacronym{miso}{MISO}{multiple-input-single-output}

\newacronym{mumiso}{MU-MISO}{multiuser multiple-input-single-output}

\newacronym{mld}{MLD}{maximum likelihood detector}
\newacronym{mrt}{MRT}{maximum ratio transmission}

\newacronym{ndris}{ND-RIS}{non-diagonal reconfigurable intelligent surfaces}

\newacronym{pls}{PLS}{physical-layer security}

\newacronym{pc}{PC}{power control}

\newacronym{qos}{QoS}{quality of service}

\newacronym{ris}{RIS}{reconfigurable intelligent surfaces}
\newacronym{re}{RE}{reflecting element}
\newacronym{rf}{RF}{radio-frequency}
\newacronym{rhs}{RHS}{right-hand side}

\newacronym{sc}{SC}{single-connected}

\newacronym{svd}{SVD}{singular value decomposition}

\newacronym{se}{SE}{spectral efficiency}
\newacronym{snr}{SNR}{signal-to-noise ratio}
\newacronym{siso}{SISO}{single-input-single-output}
\newacronym{sinr}{SINR}{signal to interference and noise ratio}
\newacronym{swipt}{SWIPT}{simultaneous wireless information and power transfer}

\newacronym{tas}{TAS}{transmit antenna selection}

\newacronym{ul}{UL}{uplink}
\newacronym{uav}{UAV}{unmanned aerial vehicle}

\newacronym{zf}{ZF}{zero forcing}

\renewenvironment{thebibliography}[1]{
  \begin{oldthebibliography}{#1}
    \setlength{\itemsep}{0.01em}
    \setlength{\parskip}{-0.12em}
}
{
  \end{oldthebibliography}
}

\begin{document}
\title{\LARGE Beyond~Diagonal~RIS for Physical-Layer Security: Low-Complexity Solution in the Low-SNR Regime}
\author{Rawan~Derbas,~
\IEEEmembership{Member,~IEEE,} and Hamad~Yahya,~\IEEEmembership{Member,~IEEE} 
\thanks{The work of Rawan Derbas and Hamad Yahya was funded by Khalifa University of Science and Technology through the Faculty Startup Grant, under Project ID: KU-INT-FSU-2026-8471000026. Rawan Derbas and Hamad Yahya are with Center for Intelligent Computing, Networks and Sensing (ICONS), Department of Computer and Information Engineering, Khalifa University, Abu Dhabi, United Arab Emirates (email: \href{mailto:rawan.derbas@ku.ac.ae}{rawan.derbas@ku.ac.ae}, \href{mailto:hamad.myahya@ku.ac.ae}{hamad.myahya@ku.ac.ae}).\\
Interested readers can find the source code for this letter at \href{https://github.com/RawanMad/BD-RIS-for-PLS-Low-SNR}{https://github.com/RawanMad/BD-RIS-for-PLS-Low-SNR}.
}}%

\maketitle
\begin{abstract}
This letter investigates \gls{bdris} for \gls{pls} in correlated wiretap channels. A multi-antenna transmitter is considered, and \gls{tas} is used to exploit transmit-side diversity with a single active \gls{rf} chain. Using the first-order secrecy rate approximation, we develop a two-stage scheme to maximize the secrecy rate in the low \gls{snr} regime. In the first stage, two \gls{tas} schemes are developed based on the effective channel upper-bound. In the second stage, an eigenvector-based \gls{bdris} scattering matrix is designed for single-connected, group-connected, and fully-connected architectures. Unlike legitimate-user-centric or eavesdropping-nulling designs, the proposed approach balances legitimate-user signal enhancement and information-leakage suppression without forcing zero leakage. Simulation results show improved secrecy performance, particularly under a strong correlation between legitimate and eavesdropping channels.
\end{abstract}

\glsresetall

\begin{IEEEkeywords}
\Gls{bdris}, low \gls{snr}, \gls{pls}, nulling.
\end{IEEEkeywords}
\glsresetall

\markboth{Draft,~Vol.~xx, No.~xx,
OCT.~2026}{DERBAS and YAHYA: Beyond~Diagonal~RIS for Physical-Layer Security: Low-Complexity Solution in Low-SNR Regime} 

\section{Introduction}
\IEEEPARstart{T}{he} broadcast nature of wireless communications makes transmitted signals accessible to unintended receivers, rendering \gls{pls} essential for next-generation wireless systems \cite{wang2022wireless}. This challenge is particularly important in power-limited networks, where low received \glspl{snr} make reliable and secure communication more challenging \cite{wang2018survey}. \Gls{ris} offer a promising solution by passively reshaping the propagation environment to enhance legitimate reception and limit information leakage \cite{di2020smart}. \Gls{bdris} further extend this capability through reconfigurable interconnections among \glspl{re}, providing greater channel-shaping flexibility than conventional \gls{dris} \cite{li2022beyond,li2025tutorial}.

Despite these advantages, \gls{bdris} secrecy design remains challenging due to its nonconvex constraints imposed by the architecture of \gls{bdris}, with many existing methods relying on iterative optimization \cite{BDRIS_PLS1,pls_bd2,pls_bd3}. Recent closed-form designs with low-complexity have been introduced to improve the quality of received signals in communication users \cite{closed1,fang2024low,yahya2024beyond}. However, these designs may simultaneously improve the quality of the received signal at both the legitimate users and eavesdroppers, especially when their channels are highly correlated. 

To address this challenge, we propose a low-complexity, closed-form \gls{bdris} design targeting secrecy rate enhancement in power-limited systems. Specifically, we develop a non-iterative, eigenvector-based design tailored to the low-\gls{snr} regime without imposing strict eavesdropper nulling. The proposed scheme is adapted to \gls{sc}, \gls{gc}, and \gls{fc} \gls{bdris} architectures. Furthermore, we exploit \gls{tas} to preserve full transmit diversity while maintaining low-complexity and power efficiency \cite{Yeoh2011MIMORelaying}. We evaluate the proposed design against passive \gls{mrt} and eavesdropper-nulling benchmarks \cite{yahya2024beyond}. Simulation results demonstrate improved secrecy performance over both benchmarks, particularly under strong correlation between the legitimate and eavesdropping channels.

\textit{Organization:}
The rest of the letter is organized
as follows. Sec. \ref{sec-model} presents the system model and problem formulation, while Sec. \ref{sec-sol} presents the proposed secure scheme.
Sec. \ref{sec-res} presents the results and discussions. Finally, Sec. \ref{sec-concl} is the conclusions.

\textit{Notations:}
The notations used throughout the paper are explained as follows. Boldface uppercase and lowercase symbols such as $\mathbf{S}$ and
$\mathbf{s}$ denote matrices and column vectors, respectively. The absolute value of a scalar is denoted by $|\cdot|$, while $(\cdot)^T$ and $(\cdot)^H$ denote the transpose and Hermitian transpose, respectively. $[\mathbf{X}]_{i,j}$ denotes the $i$-th row and $j$-th column entry of $\mathbf{X}$. $\|\cdot\|_2$ the Euclidean norm. The identity matrix is denoted by $\mathbf{I}$. Also, $\text{diag}(x)$ is a diagonal matrix with the main diagonal
being $x$, and $\text{blkdiag}(X_1,X_2,\dots,X_G)$ is a $G \times G$ block matrix
with diagonal blocks
being $X_1,X_2,\dots,X_G$. A circularly symmetric complex Gaussian
random variable with zero mean and variance $\sigma^2$ is denoted by $\mathcal{CN}(0,\sigma^2)$.
\begin{figure}[h]
\centering
\includegraphics[
 width=3.6in,
 trim=3.5cm 2.5cm 3cm 2.8cm,
 clip
]{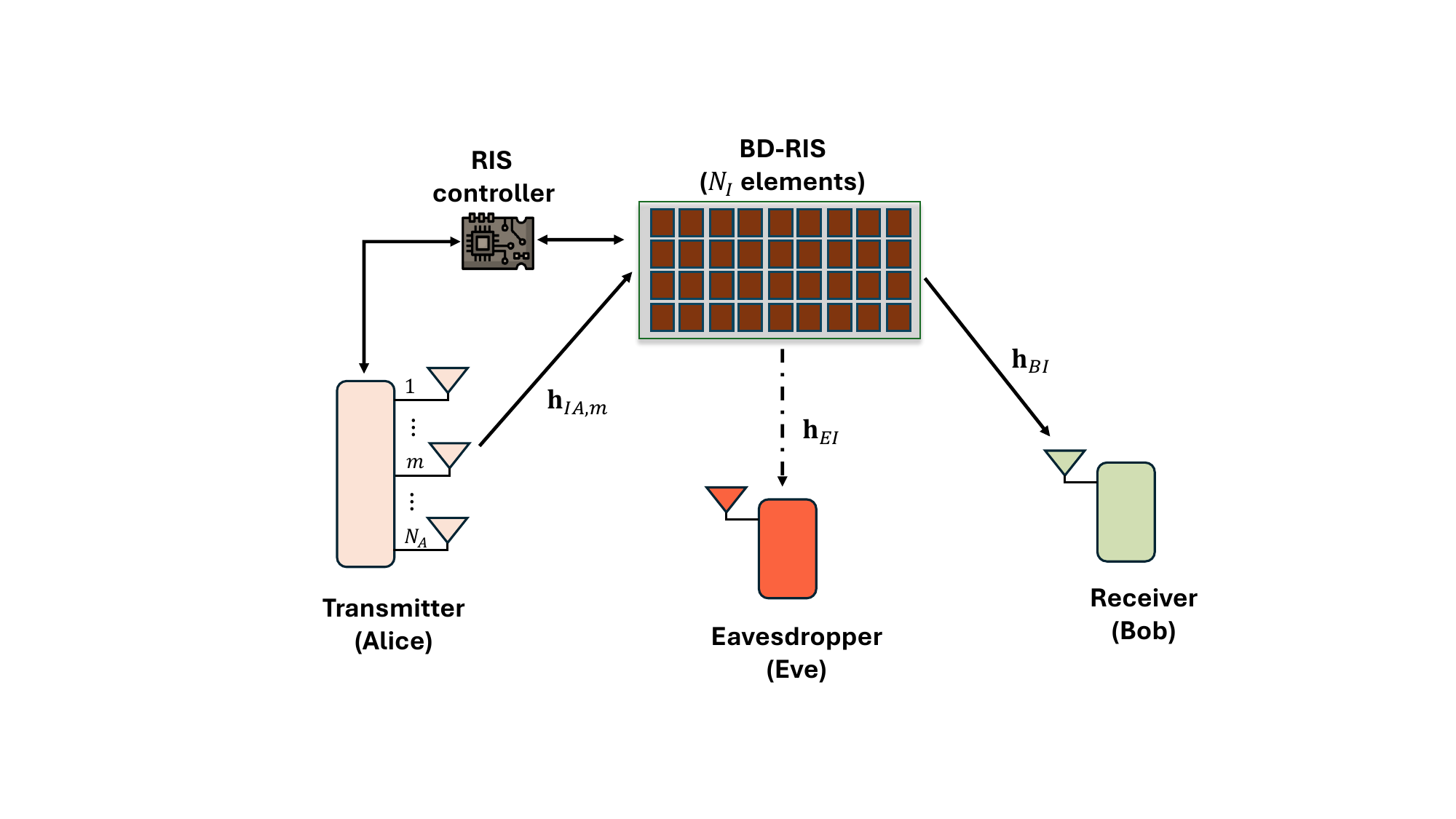}
\caption{Proposed System Model.}
\label{fig:system_model}
\end{figure}
\section{System Model and Problem Formulation}\label{sec-model}

\subsection{MISO-TAS BD-RIS Wiretap Model}
In this paper, we consider a \gls{bdris}-assisted secure \gls{miso} system, in which consists of a multi-antenna transmitter (Alice), two single-antenna receivers, the legitimate user (Bob), the eavesdropper (Eve), and an $N_I$- elements \gls{bdris}, as shown in Fig. \ref{fig:system_model}.
The channels from \gls{bdris} to Bob, and from \gls{bdris} to Eve are denoted by $\mathbf{h}_{BI}\in\mathbb{C}^{1\times N_I}$, $\mathbf{h}_{EI}\in\mathbb{C}^{1\times N_I}$, respectively. 
Also, The channel from Alice's $m$-th antenna to the \gls{bdris} is denoted as $\mathbf{h}_{IA,m}\in\mathbb{C}^{N_I\times 1}$. Thus, when antenna $m$
is selected, the received signals at Bob and Eve are respectively given by
\begin{equation}
 \label{y_b}
 y_B=\sqrt{P_t}( \mathbf{h}_{BI}\mathbf{\Theta}\mathbf{h}_{IA,m})s+n_B
\end{equation}
\begin{equation}
 \label{y_e}
 y_E=\sqrt{P_t}( \mathbf{h}_{EI}\mathbf{\Theta}\mathbf{h}_{IA,m})s+n_E
\end{equation}
where $s$ denotes the information symbol intended for Bob with $\mathbb{E}[|x|^2]=1$, $P_t$ is the transmit power, and $\mathbf{\Theta}\in \mathrm{C}^{N_I \times N_I}$ is the \gls{bdris} scattering matrix. 
Also, $n_k \sim \mathcal{CN}(0,\sigma_k^2)$, $k \in \{B,E\}$.
Thus, the received \glspl{snr} at Bob and Eve are given as
\begin{equation}
\label{SNR_B}
 \gamma_{B,m}(\boldsymbol{\Theta})=\bar{\gamma}_B| \mathbf{h}_{BI}\mathbf{\Theta}\mathbf{h}_{IA,m}|^2
\end{equation}
\begin{equation}
\label{SNR_E}
 \gamma_{E,m}(\boldsymbol{\Theta})=\bar{\gamma}_E| \mathbf{h}_{EI}\mathbf{\Theta}\mathbf{h}_{IA,m}|^2
\end{equation}
where $\bar{\gamma}_B=\frac{P_t}{\sigma_B^2}$, and $\bar{\gamma}_E=\frac{P_t}{\sigma_E^2}$. Consequently, the instantaneous secrecy rate can be written as
\begin{equation}
\label{secrecy_rate}
R_{s,m}(\boldsymbol{\Theta})
\!=\!
\Big[
\underbrace{\log_2\big(1 + \gamma_{B,m}(\boldsymbol{\Theta})\big)}_{\triangleq R_{B,m}}
\!-\!
\underbrace{\log_2\big(1 + \gamma_{E,m}(\boldsymbol{\Theta}))\big)}_{\triangleq R_{E,m}}
\Big]^+.
\end{equation}
In the low-\gls{snr}regime, the
secrecy rate admits a first-order approximation. Specifically,
using the approximation $\log_2(1+x) \approx x\log_2(e)$ for
$x \ll 1$ \cite[Eq.~(5.15)]{tse2005fundamentals}, we obtain
\begin{equation}
\label{r_exp}
    \log_2\big(1 + \gamma_{K,m}(\boldsymbol{\Theta})\big)=\gamma_{K,m}(\boldsymbol{\Theta})\log_2(e)
\end{equation}
where $K \in\{B,E\}$. Substituting \eqref{r_exp} into \eqref{secrecy_rate}, the secrecy
rate can be written as
\begin{equation}
\label{secrecy_rate2}
R_{s,m}(\boldsymbol{\Theta})
\!=\!\log_2(e)\!
\Big[\!
\bar{\gamma}_B\!\left|\mathbf{h}_{BI}\mathbf{\Theta}\mathbf{h}_{IA,m}\right|^2
-
\bar{\gamma}_E\!\left|\mathbf{h}_{EI}\mathbf{\Theta}\mathbf{h}_{IA,m}\right|^2
\!\Big]^+.
\end{equation}
\subsection{BD-RIS Architecture Constraints} 
For purely reactive impedance and reciprocal \gls{bdris} implementations, the scattering
matrix, $\boldsymbol{\Theta}$, is symmetric and unitary within the permitted connection
structure. Depending on the circuit topology for the \gls{bdris} reconfigurable impedance matrix, three architectures are considered \cite{li2022beyond}:
\begin{itemize}
    \item \gls{sc} \gls{bdris}: Also known as the conventional \gls{dris}. In this architecture, each \gls{re} is controlled independently, this implies a diagonal scattering matrix with 
unit-modulus entries given, respectively, by
\begin{equation}
    \mathcal{S}_{\text{SC}_1}
    = \{\mathbf{\Theta} : [\mathbf{\Theta}]_{i,j} = 0,\ \forall\, i \neq j\}
\end{equation}
\begin{equation}
    \mathcal{S}_{\text{SC}_2}
    = \{\mathbf{\Theta} : \big|[\mathbf{\Theta}]_{i,j}\big| = 1,\ \forall\, i = j\}
\end{equation}
\item \gls{fc} \gls{bdris} architecture is realized by interconnecting each \gls{re} with all other \glspl{re} through a reciprocal reconfigurable impedance network. Thus, the corresponding full scattering matrix, denoted as $\mathbf{\Theta}\in \mathbb{C}^{N_I\times N_I}$, is symmetric and unitary, i.e., 
\begin{equation}
   \mathcal{S}_{\text{FC}_1}=\big\{\mathbf{\Theta}:\mathbf{\Theta}=\mathbf{\Theta}^T\big\}
\end{equation}
 \begin{equation}
  \mathcal{S}_{\text{FC}_2}= \big\{\mathbf{\Theta}:\mathbf{\Theta}\mathbf{\Theta}^H=\mathbf{I}\big\}
  \end{equation}
 
\item \gls{gc} \gls{bdris}: 
For the group connected architecture, the $N_I$ elements are
partitioned into $G=N_I/N_G$ groups of $N_G$
elements each. The scattering matrix has the
block-diagonal structure, with each block
being symmetric and unitary, thus the following constraints shall satisfy
\begin{multline}\label{eq:GC_structure}
\mathcal{S}_{\text{GC}_1}=\big\{\boldsymbol{\Theta}:\boldsymbol{\Theta}
=
\operatorname{blkdiag}
(\boldsymbol{\Theta}_1,\ldots,\boldsymbol{\Theta}_G),\\ 
\boldsymbol{\Theta}_g=\boldsymbol{\Theta}_g^T,\forall g\big\}
\end{multline}
and 
 \begin{equation}
  \mathcal{S}_{\text{GC}_2}= \big\{\mathbf{\Theta}:\mathbf{\Theta}_g\mathbf{\Theta}_g^H=\mathbf{I}\big\}
  \end{equation}
  where $\mathbf{\Theta}_g \in \mathbf{C}^{N_g \times N_g}$, $g\in\{1,2,\dots,G\}$. We can see that the fully connected can be obtained if we consider $G=1$, and the single connected when $G=N_I$.

\end{itemize}

Based on the above, the secrecy rate maximization
problem for the considered \gls{bdris} architectures is
formulated as
\begin{subequations}\label{eq:opt_SR}
\begin{align}
(P1):\quad \max_{m,\boldsymbol{\Theta}}\quad &
R_{s,m}(\boldsymbol{\Theta})
\label{prob:SR_obj}\\
\mathrm{s.t.}\quad &
\boldsymbol{\Theta}\in\mathcal{S}_{c_1}
\label{prob:SR_sym}\\
&
\boldsymbol{\Theta}\in\mathcal{S}_{c_2}
\label{prob:SR_unit}\\
&
m\in\mathcal{N}_A
\end{align}
\end{subequations}
where $c\in\{\mathrm{SC},\mathrm{GC},\mathrm{FC}\}$,
and $\mathcal{N}_A=\{1,\ldots,N_A\}$.

\section{Proposed Scheme}\label{sec-sol}

Solving \eqref{eq:opt_SR} directly is challenging
due to the discrete antenna-selection variable and the nonconvex \gls{bdris} constraints. Therefore, we propose a low-complexity two-stage design that first selects the transmit antenna and subsequently configures the \gls{bdris} for the selected antenna.

\subsection{Stage 1: Transmit Antenna Selection}

To select the transmit antenna independently of
the \gls{bdris} configuration, we use the
Cauchy--Schwarz inequality to upper-bound
the effective channel gains as
\begin{equation}
\label{eq:CS_bound}
|\mathbf{h}_{kI}\boldsymbol{\Theta}
\mathbf{h}_{IA,m}|^2
\leq
\|\mathbf{h}_{kI}\|_2^2
\|\mathbf{h}_{IA,m}\|_2^2,
\quad k\in\{B,E\}.
\end{equation}
Since $\|\mathbf{h}_{kI}\|_2^2$ is independent
of $m$, two antenna-selection criteria are considered:
\begin{enumerate}
    \item Bob-based selection, which maximizes the upper bound on Bob's channel gain.
    \item Eve-based selection, which minimizes the upper bound on Eve's channel gain.
\end{enumerate}
The corresponding antenna indices are given by
\begin{align}
m_B^\star
&=\arg\max_{m\in\mathcal{N}_A}
\|\mathbf{h}_{IA,m}\|_2^2
\label{eq:tas_bob}\\
m_E^\star
&=\arg\min_{m\in\mathcal{N}_A}
\|\mathbf{h}_{IA,m}\|_2^2
\label{eq:tas_eve}
\end{align}
Following antenna selection, the corresponding
$\mathbf{h}_{IA,m}$ is used to design the
\gls{bdris} scattering matrix in the second stage.

\subsection{Stage 2: Low-SNR Secure BD-RIS Design}
For a fixed antenna $m^{\star}\in\{m_E^{\star},m_B^{\star}\}$, expressing the squared magnitudes in \eqref{secrecy_rate2}
as quadratic forms yields 
\begin{equation}
\label{secrecy_rate3}
R_{s}(\boldsymbol{\Theta})
=\Big[\log_2(e)\big(
\mathbf{h}_{IA,m^{\star}}^H\mathbf{\Theta}^H \mathbf{Q}
\mathbf{\Theta}\mathbf{h}_{IA,m^{\star}}\big)
\Big]^+
\end{equation}
where $\mathbf{Q}=\bar{\gamma}_B\mathbf{h}_{BI}^H\mathbf{h}_{BI}
-
\bar{\gamma}_E\mathbf{h}_{EI}^H\mathbf{h}_{EI}$ where $\mathbf{Q} \in \mathrm{C}^{N_I \times N_I}$.
Let $\mathbf{x}=\mathbf{\Theta}\mathbf{h}_{IA,m^{\star}}$, the \gls{bdris} design problem in \eqref{eq:opt_SR} can be reformulated considering low-\gls{snr} regime as
\begin{subequations}\label{eq:opt_SR2}
\begin{align}
(P2):\max_{\boldsymbol{\Theta}} \quad &
\mathbf{x}^H \mathbf{Q}\mathbf{x}
\label{prob:SR_obj}\\
\text{s.t.}\quad &
\eqref{prob:SR_sym}, \eqref{prob:SR_unit}
\end{align}
\end{subequations}
To solve $(P2)$, we first relax the symmetric and unitary constraints. We, then invoke the Rayleigh theorem \cite[Theorem~4.2.2]{horn2012matrix} to derive an upper bound on the objective and identify the alignment condition required to attain it.

Since $\bar{\gamma}_B$ and $\bar{\gamma}_E$ are real,
$\mathbf{Q}$ is Hermitian:
\begin{equation}
\mathbf{Q}^H
=
\bar{\gamma}_B\mathbf{h}_{BI}^H\mathbf{h}_{BI}
-
\bar{\gamma}_E\mathbf{h}_{EI}^H\mathbf{h}_{EI}
=
\mathbf{Q}.
\end{equation}
Since $\mathbf{Q}$ is Hermitian with $R_{\mathbf{Q}}=\text{rank}(\mathbf{Q})\leq \min\{2,N_I\}$, its eigenvalue
decomposition is given by
$\mathbf{Q}=\mathbf{U}\boldsymbol{\Lambda}\mathbf{U}^H$,
where $\mathbf{U}=[\mathbf{u}_1,\ldots,\mathbf{u}_{N_I}]\in \mathrm{C}^{N_I \times N_I}$
contains the orthonormal eigenvectors of $\mathbf{Q}$,
and
$\boldsymbol{\Lambda}=\operatorname{diag}
(\lambda_1,\ldots,\lambda_{N_I})\in \mathrm{C}^{N_I \times N_I}$
contains the corresponding real-valued eigenvalues, arranged in descending order, with $R_{\mathbf{Q}}$ nonzero eigenvalues and $N_I-R_{\mathbf{Q}}$ zero eigenvalues.
Thus, $\lambda_{\max}(\mathbf{Q})=\lambda_1$,
and its associated unit-norm eigenvector is
$\mathbf{u}_{\max}=\mathbf{u}_1$.
Moreover, for every feasible $\boldsymbol{\Theta}\in \mathcal{S}_{c_2}$, $\mathbf{x}$ need to be a unit vector. Thus, we define
\begin{equation}
\label{u_IA_def}
\mathbf{u}_{IA,m^\star}
=
\frac{\mathbf{h}_{IA,m^\star}}
{\|\mathbf{h}_{IA,m^\star}\|_2},
\qquad
\mathbf{y}
=
\boldsymbol{\Theta}\mathbf{u}_{IA,m^\star}.
\end{equation}
Then
$\mathbf{x}=\|\mathbf{h}_{IA,m^\star}\|_2\mathbf{y}$
and $\|\mathbf{y}\|_2=1$ for every feasible
$\boldsymbol{\Theta}$. The Rayleigh theorem
therefore gives
\begin{equation}
\mathbf{y}^H\mathbf{Q}\mathbf{y}
\leq
\lambda_{\max}(\mathbf{Q}).
\end{equation}
Consequently,
\begin{equation}
\label{bound_corrected}
\max_{\boldsymbol{\Theta} \in \mathcal{S}_{c_2}}
\mathbf{x}^H\mathbf{Q}\mathbf{x}
\leq
\lambda_{\max}(\mathbf{Q})
\|\mathbf{h}_{IA,m^\star}\|_2^2
\end{equation}
The equality in \eqref{bound_corrected} holds if and only if $
\mathbf{Q}\mathbf{y}
=
\lambda_{\max}(\mathbf{Q})\mathbf{y}.
$
Since $\mathbf{u}_{\max}$ is the eigenvector
associated with $\lambda_{\max}(\mathbf{Q})$, a sufficient condition for a feasible scattering matrix
to attain the bound in \eqref{bound_corrected} is
\begin{equation}
\label{eq:stage2_alignment}
\boldsymbol{\Theta}\mathbf{u}_{IA,m^\star}
=
\mathbf{u}_{\max}.
\end{equation}

In order to satisfy the above alignment \begin{equation}
\label{theta_B}
\boldsymbol{\Theta}
=
\mathbf{u}_{\max}\mathbf{u}_{IA,m^\star}^H,
\end{equation}
However, satisfying the alignment equation alone does
not guarantee feasibility.
To establish attainment of the upper bound in $(P2)$,
a scattering matrix $\boldsymbol{\Theta}^{\star}$
must satisfy the symmetry and unitary constraints. This construction depends on the constraints imposed by the specific BD-RIS architectures as follows.

\subsubsection{Single-Connected Architecture}

The \gls{sc} architecture restricts the scattering matrix
to a diagonal form with unit-modulus entries. Thus, the scattering coefficients are constructed as
\begin{equation}
[\boldsymbol{\Theta}_{\mathrm{SC}}^{\star}]_{i,j}
=
\begin{cases}
\mathcal{S}_{\mathrm{unim}}\!\left(
[\mathbf{u}_{\max}]_i[\mathbf{u}_{IA,m^{\star}}]_j^{*}
\right),
& i=j\\
0,
& i\neq j
\end{cases}
\label{eq:SC_phase_solution}
\end{equation}
where $
\mathcal{S}_{\mathrm{unim}}(z)
=\frac{z}{|z|}$.
\subsubsection{Group-Connected and Fully Connected  Architectures}
Starting from the unconstrained scattering matrix 
\begin{equation}
 {\boldsymbol{\Theta}}
=
\mathbf{u}_{\max}\mathbf{u}_{IA,m^{\star}}^{H}
\end{equation}
the group-$g$ target block is extracted as
\begin{equation}
\label{GC_Theta1}
 {\boldsymbol{\Theta}}_{g}
=
[ {\boldsymbol{\Theta}}]_{\mathcal{I}_g,\mathcal{I}_g}
=
[\mathbf{u}_{\max}\mathbf{u}_{IA,m^{\star}}^{H}]_{\mathcal{I}_g,\mathcal{I}_g}
\end{equation}
where
$\mathcal{I}_g=\{(g-1)N_G+1,\ldots,gN_G\}$.
Since $ {\boldsymbol{\Theta}}_{g}$ does not, in general, satisfy the symmetric unitary constraints 
imposed by the \gls{bdris} architecture (i.e., \eqref{prob:SR_sym}, \eqref{prob:SR_unit}), we employ the symmetric unitary projection proposed in \cite{fang2024low}.
First, the symmetric part is obtained as $\mathcal{S}_{\text{sym}}(\boldsymbol{\Theta}) = \tfrac{1}{2}\big(\boldsymbol{\Theta} + \boldsymbol{\Theta}^{T}\big)$. Next, let  $R =\text{rank}(\mathcal{S}_{\text{sym}}(\boldsymbol{\Theta}))$ and compute \gls{svd} of $\mathcal{S}_{\text{sym}}(\boldsymbol{\Theta})$  by evaluating $\mathcal{S}_{\text{sym}}(\boldsymbol{\Theta})= \mathbf{W}\boldsymbol{\Sigma}\mathbf{V}^{H}$. Here, $\mathbf{W} \in \mathrm{C}^{N_I \times N_I}$ and $\mathbf{V} \in \mathrm{C}^{N_I \times N_I}$
are the unitary left and right singular vectors and $\boldsymbol{\Sigma} \in \mathrm{R}^{N_I \times N_I}$ contains
the singular values ordered in a descending order. Partitioning $\mathbf{W}=[\mathbf{W}_R,\mathbf{W}_{N_I-R}]$ and $\mathbf{V}=[\mathbf{V}_R,\mathbf{V}_{N_I-R}]$, the symmetric-unitary projection is then given by
\begin{equation}
\label{unit_sym}
\mathcal{S}_{\text{uni}}(\mathcal{S}_{\text{sym}}(\boldsymbol{\Theta})) = \widehat{\mathbf{W}}\mathbf{V}^{H}
\end{equation}
where $\widehat{\mathbf{W}}=[\mathbf{W}_R, \mathbf{V}_{N_I-R}^*]$.\\The \gls{gc} scattering matrix is then constructed as
\begin{equation}
\label{GC_Theta2}
\boldsymbol{\Theta}_{\mathrm{GC}}^{\star}
=
\operatorname{blkdiag}
\left(
\boldsymbol{\Theta}_{1}^{\star},
\ldots,\boldsymbol{\Theta}_{g}^{\star},\dots,
\boldsymbol{\Theta}_{G}^{\star}
\right),
\end{equation}
where $\boldsymbol{\Theta}_{g}^{\star} \in \mathcal{S}_{\text{symuni}}$.

For $G=1$, the \gls{gc} construction contains a single
$N_I\times N_I$ symmetric-unitary block and therefore reduces
to the \gls{fc} architecture. The pseudo-code for the proposed algorithm is provided at the top of the next page.
\par \textit{Computational Complexity:}
With $N_I=GN_g$, antenna selection requires
$\mathcal{O}(N_A GN_g)$ operations. The second stage
requires $\mathcal{O}(G^2N_g^2)$ operations to form
$\mathbf{Q}$, $\mathcal{O}(G^3N_g^3)$ operations for
its full eigen decomposition, and
$\mathcal{O}(GN_g^3)$ operations for the group-wise
SVD-based symmetric unitary projections. Hence,
the overall complexity is
$\mathcal{O}(N_A GN_g+G^3N_g^3)$.
\NoCaptionOfAlgo
\begin{algorithm}[h]
\caption{\textbf{Algorithm:} Eigenvector-Based \gls{bdris} Design}
\label{alg:FC_secrecy_design}

\KwInput{ $G$, $\mathbf{h}_{IA,m^{\star}}\in\mathbb{C}^{N_I\times 1}
       $,
         $\mathbf{h}_{BI},\mathbf{h}_{EI}
         \in\mathbb{C}^{1\times N_I}$,
         $\bar{\gamma}_B,\bar{\gamma}_E$} 
\KwOutput{$\boldsymbol{\Theta^{\star}}$}

$\mathbf{Q}
       =\bar{\gamma}_B\mathbf{h}_{BI}^{H}\mathbf{h}_{BI}
       -\bar{\gamma}_E\mathbf{h}_{EI}^{H}\mathbf{h}_{EI}$\;

$\mathbf{u}_{IA,m^{\star}}
       =\mathbf{h}_{IA,m^{\star}}/\|\mathbf{h}_{IA,m^{\star}}\|_2$\;

Compute the unit-norm eigenvector
       $\mathbf{u}_{\max}$ associated with
       $\lambda_{\max}(\mathbf{Q})$\;

$\mathbf{\Theta}
       =\mathbf{u}_{\max}\mathbf{u}_{IA,m^{\star}}^{H}$\;
\eIf{$G=N_I$}{
    Construct $\boldsymbol{\Theta}^{\star}$
    using \eqref{eq:SC_phase_solution}\;
  }{
    \For{$g=1,\ldots,G$}{
Construct $\boldsymbol{\Theta}_g$ using \eqref{GC_Theta1}\;

$\mathcal{S}_{\text{sym}}(\boldsymbol{\Theta}_g)
       =\frac{1}{2}(\mathbf{\Theta}_g+\mathbf{\Theta}^{T}_g)$\;

Apply the symmetric unitary projection using \eqref{unit_sym}\;

 Construct $\boldsymbol{\Theta}_{\text{GC}}^{\star}$
    using \eqref{GC_Theta2}\;
    }
}
\Return $\boldsymbol{\Theta^{\star}}$\;
\end{algorithm}

\section{Simulation results and discussions}\label{sec-res}
In this section, we evaluate the performance of the proposed system using Monte-Carlo simulation. Without loss of generality, we assume the simulation parameters are as follows. The locations of Alice, Bob, Eve, and the \gls{bdris} are given by $(0,0)$, $(150,0)$, $(145,0)$, and $(145,5)$, respectively. The path loss exponents for all links is $\delta={2.2}$. The path loss at the reference distance of $1~\mathrm{m}$ is assumed to be $\mathcal{L}_0=-30~\mathrm{dB}$. The distance-dependent
path loss is modeled as $\mathcal{L}_{ij}(d_{ij}) = \mathcal{L}_0 d_{ij}^{-\delta_{ij}}$, where $ij \in \{IA,BI,EI\}$. The transmit power is set to $P_t=15$ dBm, and $\sigma_k^2=-80$~dBm, $k \in \{B,E\}$  and is assumed to be the same for 
both Bob and Eve.
All channels are assumed to experience Rayleigh fading. 
The simulation parameters are selected to investigate
the low-SNR regime, in which the effective received SNRs
at both Bob and Eve remain small.
  Without loss of generality, we adopt a simplified spatial correlation model inspired by \cite{Zhou_RIS_PKG_2025,zhang2017key}, where the eavesdropping and legitimate channels are partially correlated due to shared scattering in the propagation environment. We only model the small-scale fading and directly sweep the correlation level, approximating the correlated channels as
\begin{equation}
\mathbf{h}_{EI}
= 
\rho\,\mathbf{h}_{BI} + \sqrt{1-\rho^2}\,\mathbf{v}
\label{eq:rho_model}
\end{equation}
where $\mathbf{v} \sim \mathcal{CN}(\mathbf{0},\sigma_{BI}^2\mathbf{I})$. Here, $\rho \in [0,1]$ indicates how strongly these channels are correlated,
with $\rho = 0$ corresponding to fully independent channels and $\rho = 1$ to fully correlated channels. 
 
 In order to study the performance, we evaluate three algorithms which are as follows:
 \begin{enumerate}
     \item Alg.~1 is a Bob-centric design that configures the \gls{bdris} scattering matrix to maximize Bob’s received power by using only the \gls{mrt} design, without explicitly accounting for Eve \cite[Eq.~(23)]{yahya2024beyond}.
     \item Alg.~2 is the \gls{bdris} secure design that starts from the \gls{mrt}-based \gls{bdris} configuration and adjusts the scattering matrix to preserve a large received power at Bob while simultaneously suppressing Eve via nulling \cite[Algorithm 1]{yahya2024beyond}.
     \item Alg.~3 is the proposed channel difference-centric design in Sec. \ref{sec-sol}.
 \end{enumerate}
  
\begin{figure}[tb]
 \centering
\includegraphics[width=1\linewidth]{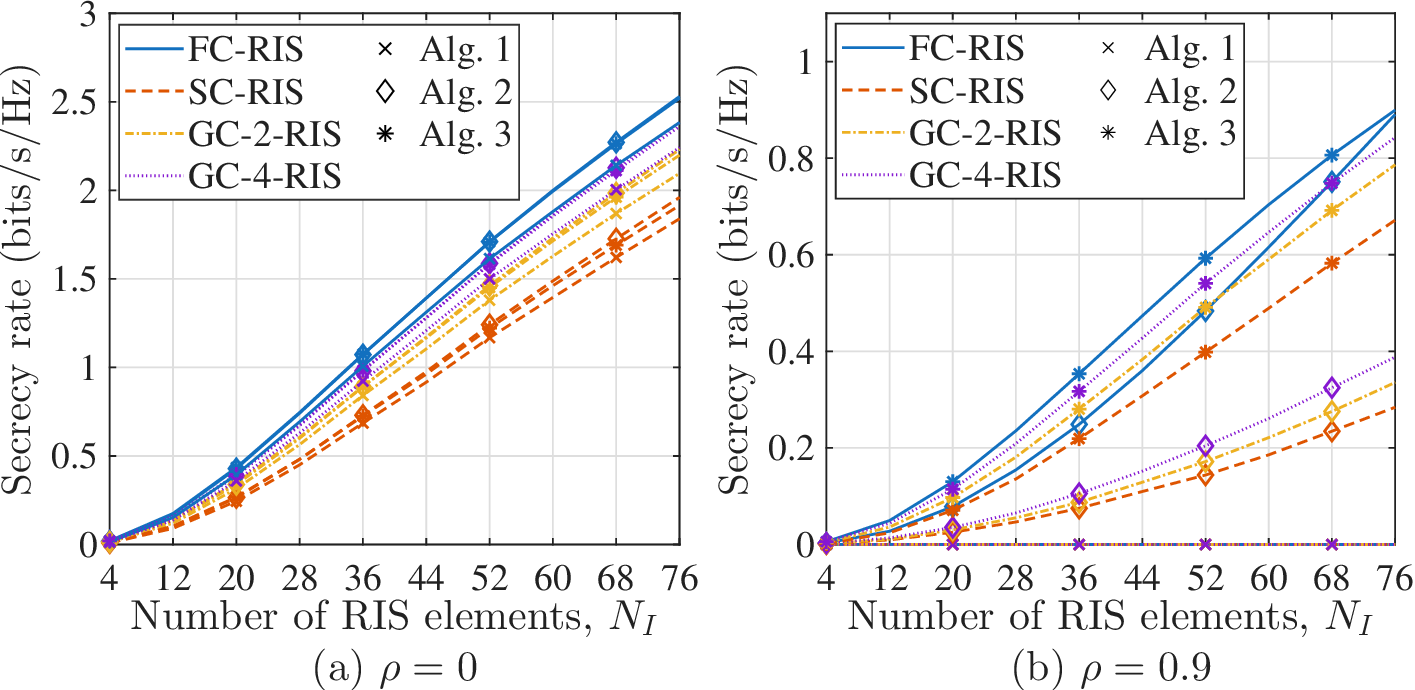}
\caption{Secrecy rate  versus $N_I$ with $N_A=4$, and $P_t=15$ $\mathrm{dBm}$ for (a) $\rho=0$ and (b) $\rho=0.9$.}
\label{fig:siso_N_rho0}
\end{figure}

We first examine \gls{miso} systems aided by different \gls{bdris} architectures considering selecting the antenna that maximizes the upper bound on Bob’s channel gain \eqref{eq:tas_bob}. In Fig.~\ref{fig:siso_N_rho0}, we compare the secrecy rate performance versus the number of \glspl{re} for independent ($\rho=0$) and strongly correlated ($\rho=0.9$) Bob-Eve channels. For $\rho=0$, the secrecy rate increases with the \gls{ris} size. The proposed
algorithm achieves secrecy rates close to those of \gls{mrt} with Eve-nulling,
particularly for larger \gls{ris} sizes. Here, Eve-nulling incurs only a
modest loss in Bob's channel gain. For $\rho=0.9$, \gls{mrt} yields approximately zero secrecy rate because
enhancing Bob's channel also strengthens Eve's link. \gls{mrt} with nulling
eliminates leakage but sacrifices substantial Bob-channel gain due to
the overlap between the channel directions. By balancing Bob-channel enhancement and Eve leakage rather than enforcing strict nulling, the proposed algorithm achieves a higher secrecy rates over the considered range.

\begin{figure}[tb]
    \centering
    \begin{minipage}[t]{0.495\linewidth}
        \centering
        \includegraphics[width=\linewidth]{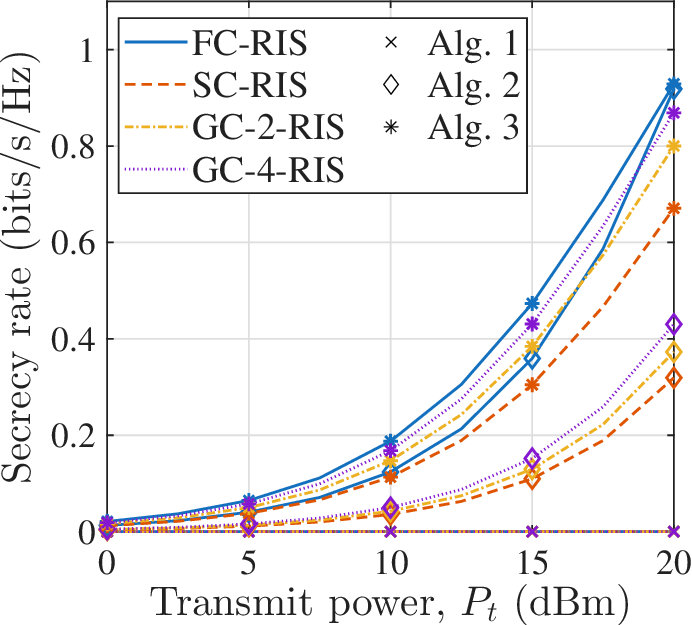}
        \caption{Secrecy rate versus $P_t$ with $N_I=44$, $N_A=4$, and $\rho=0.9$.}
        \label{fig:miso_pt}
    \end{minipage}
    \hfill
    \begin{minipage}[t]{0.475\linewidth}
        \centering
      \includegraphics[
        width=\linewidth]{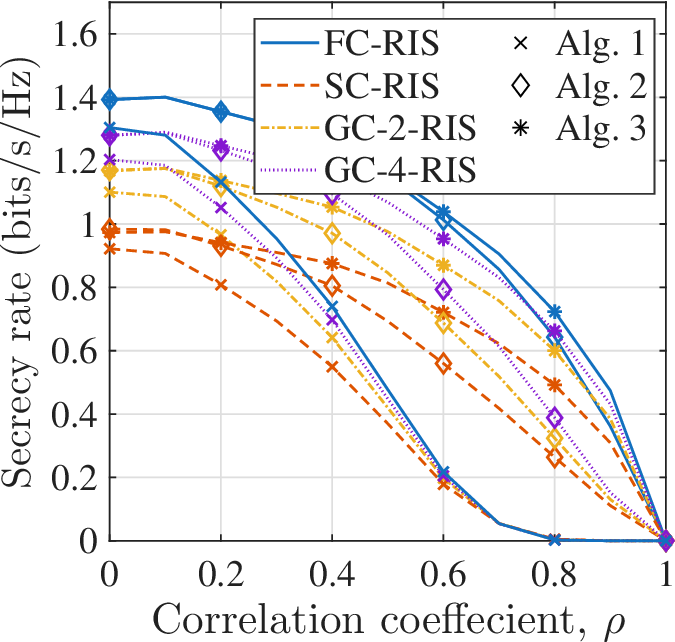}
        \caption{Secrecy rate versus $\rho$ with $N_I=44$, $N_A=4$,
        and $P_t=15\,\mathrm{dBm}$.}
        \label{fig:miso_rho}
    \end{minipage}
\end{figure}

Fig.~\ref{fig:miso_pt} shows that, at $\rho=0.9$, increasing
transmit power improves the secrecy rates of the proposed design and
\gls{mrt} with nulling, whereas \gls{mrt} remains near zero. The proposed design
achieves a better secrecy tradeoff by allowing residual leakage when
offset by greater Bob-channel gain, although its advantage over nulling
is relatively small for \gls{fc}. In both figures, the performance ordering
is \gls{fc}, \gls{gc}-4, \gls{gc}-2, and \gls{sc}.
\par Fig.~\ref{fig:miso_rho} shows the impact of Bob-Eve channel
correlation with $N_I=44$ and $P_t=15$~dBm. At low correlation, the
proposed design and \gls{mrt} with nulling achieve comparable secrecy rates.
As $\rho$ increases, \gls{mrt} approaches zero secrecy rate, while the
proposed design outperforms \gls{mrt} with nulling, since strict Eve-nulling
sacrifices useful Bob-channel gain when the channel directions strongly
overlap.

%
\begin{figure}[tb]
 \centering
\includegraphics[width=1\linewidth]{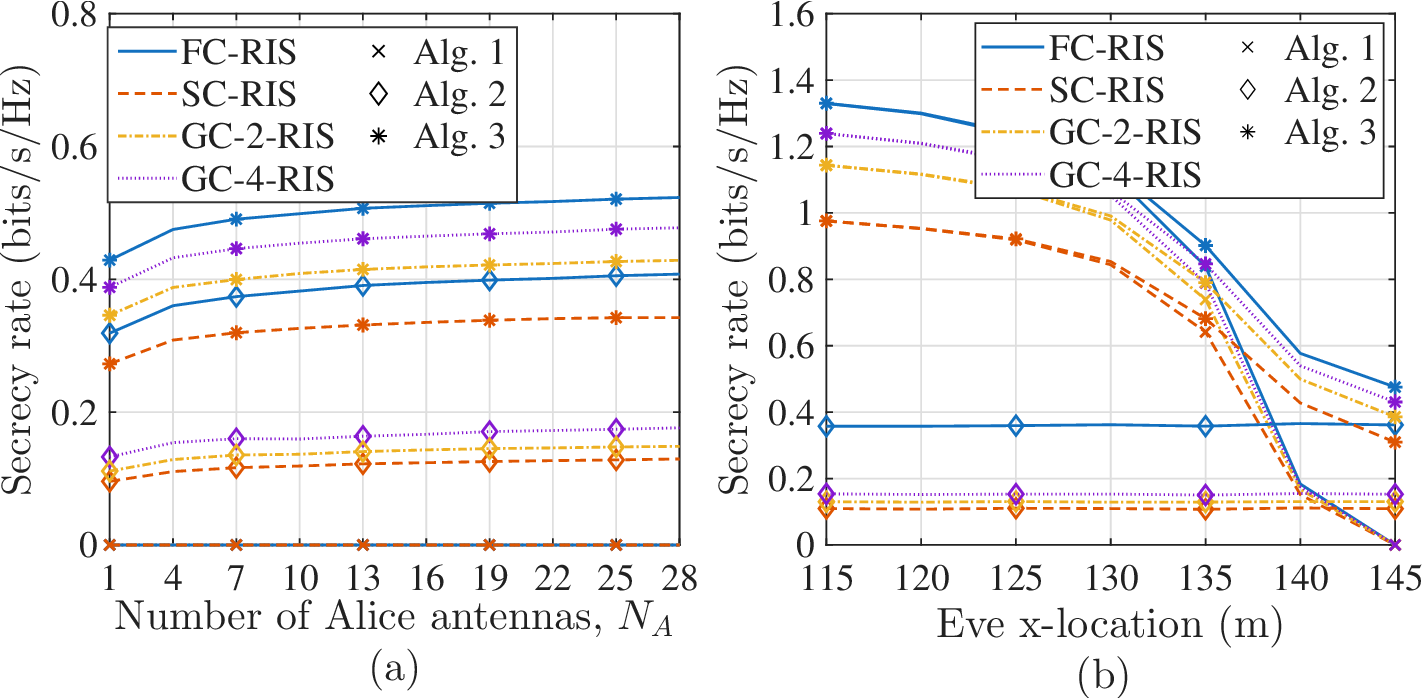}
\caption{Secrecy rate with $N_I=44$, $\rho=0.9$ and $P_t=15$ dBm versus the (a) Number of Alice antennas (b) Eve x-location.}
\label{fig:Miso_NA_rho9}
\end{figure}

Fig.~\ref{fig:Miso_NA_rho9} (a) illustrates the impact of
the number of Alice antennas, $N_A$, under strong
Bob-Eve channel correlation, $\rho=0.9$. As $N_A$ increases, the
proposed algorithm achieves an increasing secrecy rate for all \gls{bdris}
architectures. In contrast, \gls{mrt} results in a nearly zero secrecy rate,
indicating that beamforming solely toward Bob can also
enhance Eve's received signal under strong channel
correlation. Although \gls{mrt} with Eve-nulling achieves a considerable secrecy rates, its performance is lower than
that of the proposed design because strict nulling
reduces the useful signal component toward Bob. Fig.~\ref{fig:Miso_NA_rho9} (b)  studies the secrecy rate versus Eve's x-coordinate. 
As Eve's x-coordinate increases, \gls{mrt} deteriorates sharply and approaches
zero secrecy rate at $x_E = 145$~m.
In contrast, Bob-Eve Difference retains a considerable secrecy rates throughout
the considered range by balancing Bob's signal enhancement against
information leakage toward Eve.
\gls{mrt}-Nulling exhibits a nearly constant secrecy rate because ideal nulling suppresses Eve’s received signal even as her distance-dependent channel gain increases. The nearly flat curve further indicates that Bob’s effective received power remains approximately unchanged over the considered positions.

\begin{figure}[t]
 \centering
\includegraphics[width=1\linewidth]{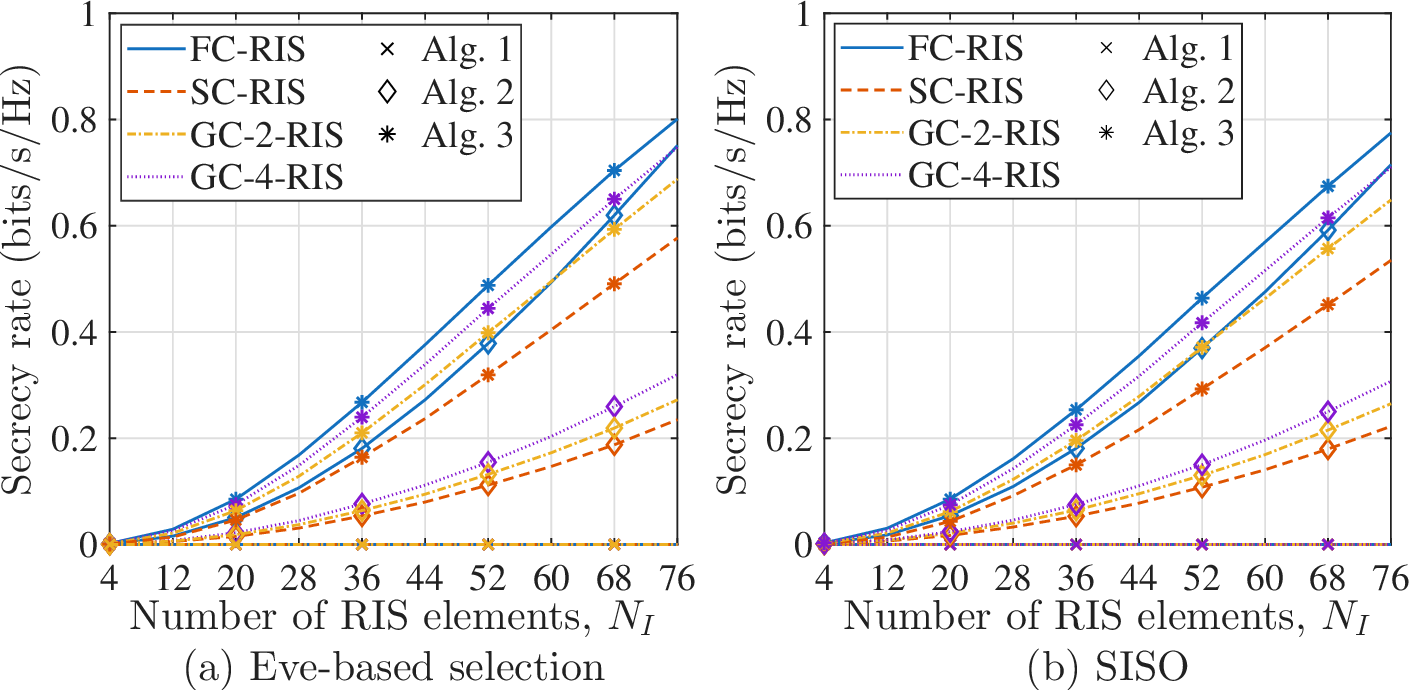}
\caption{Secrecy rate versus $N_I$ with $P_t=15$~dBm, and $N_I=44$. for (a) \gls{miso} with Eve-based selection and (b) \gls{siso}. }
\label{fig:eve_tas_siso}
\end{figure}

Fig.~\ref{fig:eve_tas_siso} compares the secrecy performance
of \gls{miso} with (a) eavesdropper-based antenna selection against
the (b) \gls{siso} baseline. A modest improvement is observed; for
example, the \gls{fc} architecture achieves an average secrecy
rate of $0.71$~bits/s/Hz at $N_I=68$, compared with
$0.67$~bits/s/Hz for \gls{siso}.

\section{conclusion}\label{sec-concl}
This letter studied low-\gls{snr} secrecy enhancement in
\gls{bdris}-assisted \gls{miso} wiretap systems.
A two-stage scheme was proposed,
comprising transmit antenna selection followed by
\gls{bdris} scattering matrix design. The
proposed algorithm maximizes the \gls{snr} advantage of Bob over Eve while satisfying the
\gls{bdris} hardware constraints.
Simulation results show that the proposed scheme
achieves secrecy rates comparable to \gls{mrt} with Eve
nulling for independent channels, while providing
substantial gains under strong Bob-Eve channel
correlation. By allowing controlled residual leakage,
it avoids the severe loss of Bob's useful signal
associated with strict nulling. Moreover, \gls{fc} achieves
the best secrecy performance, followed by \gls{gc}-4,
\gls{gc}-2, and \gls{sc}.
Future work will consider imperfect \gls{csi}, direct links,
multiple eavesdroppers, and robust \gls{mimo} extensions.
\bibliographystyle{IEEEtran}
\bibliography{bibliography.bib}
\end{document}